\documentclass[conference]{IEEEtran}
\IEEEoverridecommandlockouts
\usepackage{cite}
\usepackage{amsmath,amssymb,amsfonts}
\usepackage{algorithmic}
\usepackage{graphicx}
\usepackage{textcomp}
\usepackage{xcolor}
\usepackage{multirow}
\def\BibTeX{{\rm B\kern-.05em{\sc i\kern-.025em b}\kern-.08em
    T\kern-.1667em\lower.7ex\hbox{E}\kern-.125emX}}
\begin{document}

\title{High-risk Factor Prediction in Lung Cancer Using Thin CT Scans: An  Attention-Enhanced Graph Convolutional Network Approach*
\thanks{This work was supported by the National Natural Science Foundation of China (No.72171226). * Corresponding authors.}
}

\author{\IEEEauthorblockN{1\textsuperscript{st} Xiaotong Fu}
\IEEEauthorblockA{\textit{Center for Applied Statistics}\\
\textit{School of Statistics} \\
\textit{Renmin University of China}\\
Beijing, China \\
xiaotongfu@ruc.edu.cn}

\and
\IEEEauthorblockN{2\textsuperscript{nd} Xiangyu Meng}
\IEEEauthorblockA{\textit{Center for Applied Statistics} \\
\textit{School of Statistics}\\
\textit{Renmin University of China}\\
Beijing, China \\
mxyxy\_2156@163.com}

\and
\IEEEauthorblockN{3\textsuperscript{rd} Jing $\text{Zhou}^{*}$}
\IEEEauthorblockA{\textit{Center for Applied Statistics}\\ \textit{School of Statistics} \\
\textit{Renmin University of China}\\
Beijing, China \\
jing.zhou@ruc.edu.cn}
\and 
\IEEEauthorblockN{\hspace{7cm}4\textsuperscript{th} Ying Ji}
\IEEEauthorblockA{\hspace{7cm}\textit{Department of Thoracic Surgery} \\
\textit{\hspace{7cm}Beijing Institute of Respiratory
Medicine}\\
\textit{\hspace{7cm}Beijing Chao-Yang Hospital}\\
\textit{\hspace{7cm}Capital Medical University}\\
\hspace{7cm}Beijing, China \\
\hspace{7cm}15675112499@163.com}
}

\maketitle

\begin{abstract}
Lung cancer, particularly in its advanced stages, remains a leading cause of death globally. Though early detection via low-dose computed tomography (CT) is promising, the identification of high-risk factors crucial for surgical mode selection remains a challenge. Addressing this, our study introduces an Attention-Enhanced Graph Convolutional Network (AE-GCN) model to classify whether there are high-risk factors in stage I lung cancer based on the preoperative CT images. This will aid surgeons in determining the optimal surgical method before the operation. Unlike previous studies that relied on 3D patch techniques to represent nodule spatial features, our method employs a GCN model to capture the spatial characteristics of pulmonary nodules. Specifically, we regard each slice of the nodule as a graph vertex, and the inherent spatial relationships between slices form the edges. Then, to enhance the expression of nodule features, we integrated both channel and spatial attention mechanisms with a pre-trained VGG model for adaptive feature extraction from pulmonary nodules. Lastly, the effectiveness of the proposed method is demonstrated using real-world data collected from the hospitals, thereby emphasizing its potential utility in the clinical practice.
\end{abstract}


\section{Introduction}

\begin{figure}[htbp]
\centering
\includegraphics[width=\columnwidth]{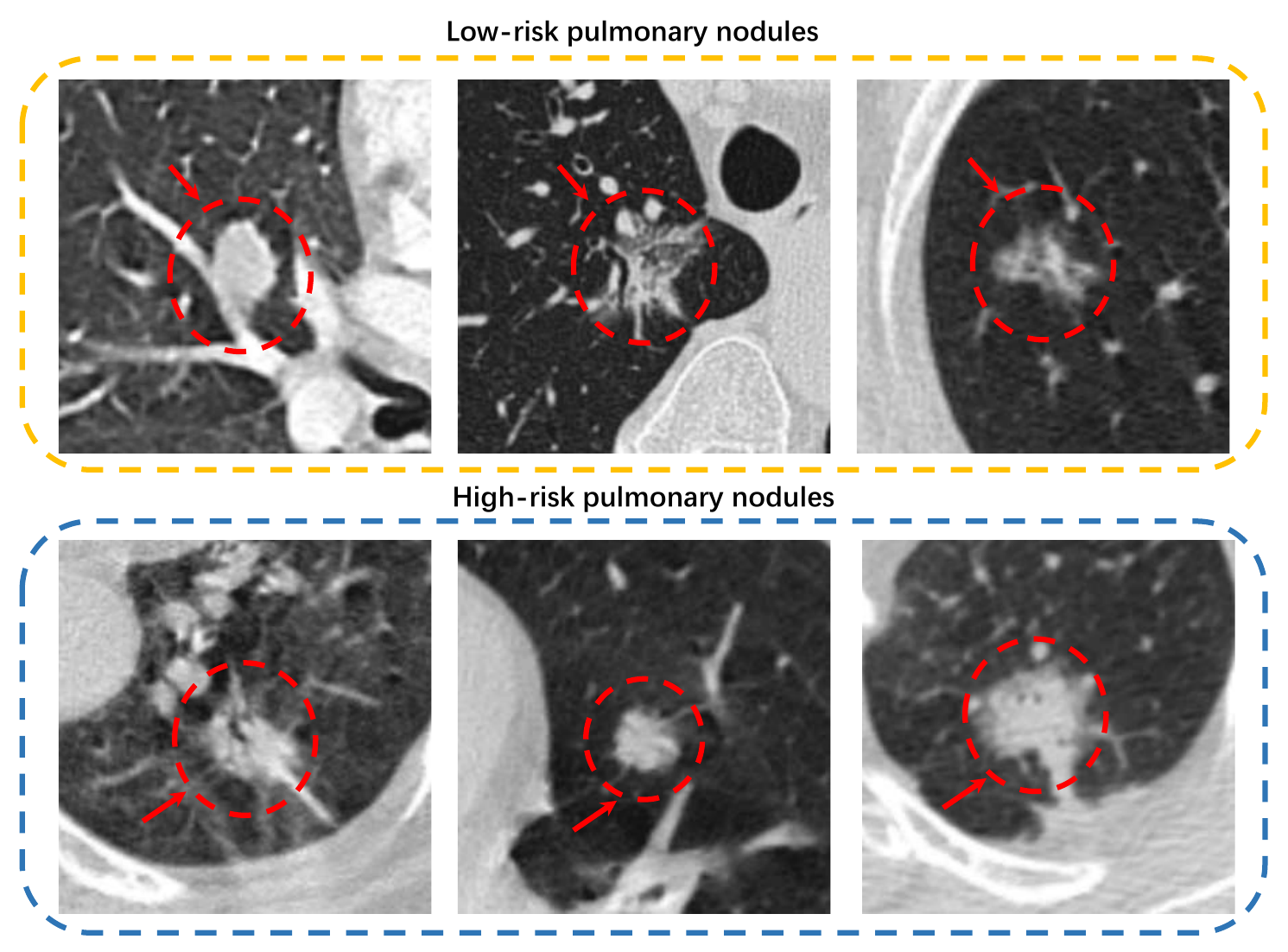}
\caption{Illustration of low-risk and high-risk nodules: the top panel shows low-risk examples, while the bottom panel presents high-risk ones.}
\label{nodules}
\end{figure}

Lung cancer is a leading cause of cancer-related deaths worldwide, posing a significant threat to human health \cite{Sung2021,Zheng2022,Siegel2022}. In China, for instance, the lack of early symptoms often results in approximately 75\% of patients being diagnosed with advanced-stage lung cancer, leading to a considerably low likelihood of cure \cite{Chinese2023}. 
%The five-year survival rate of lung cancer patients is only 19.7\%\cite{bib4}. Early detection and treatment of lung cancer can effectively improve patient survival rate (??). The early manifestation of lung cancer is pulmonary nodules, which are focal circular dense shadows with a diameter of 3mm to 3cm. 
%Lung adenocarcinoma (LUAD), as the most common type of lung cancer, represents approximately 50\% of all lung cancer patients \cite{Bray2020,Succony2021}. %During the progression of the tumor, malignant pulmonary nodules may develop into invasive lung adenocarcinoma. In the "Histological Classification of Lung Oncology" released by the World Health Organization in 2021, lung adenocarcinoma is divided into four pathological types: Atypical Adenomatous Hyperplasia (AAH), Adenocarcinoma In Situ (AIS), Minimally Invasive Adenocarcinoma (MIA), and Invasive Adenocarcinoma (IA)\cite{bib7}. 
In recent decades, with the popularization of low-dose CT screening technology, the majority of lung cancers are detected at an early stage (e.g., stage I). For those early-stage lung cancers, subpulmonary resection has demonstrated comparable outcomes to lobectomy, while preserving a greater amount of healthy lung tissue for patients  \cite{Nitadori2013}. %divided into high-risk pathological factors such as micropapillary, solid, complex glandular patterns, and vascular tumor thrombus may result in a higher rate of recurrence \cite{Su2020}. 
%Invasive adenocarcinoma can be divided into pathological subtypes such as adherent type, acinar type, papillary type, solid type, and micropapillary type. Different pathological subtypes have different surgical methods, and the prognosis also varies greatly\cite{bib8}. For some patients with low malignancy characteristics, segmentectomy is an acceptable surgery with the additional advantage of preserving lung function\cite{bib9}\cite{bib10}\cite{bib11}. For patients with high-risk pathological factors such as micropapillary type, solid type, and spread through air spaces, lobectomy has a better prognosis\cite{bib12}, Therefore, accurately identifying these high-risk pathological factors for lung cancer is very important. 
However, several studies have demonstrated that using the subpulmonary approach for patients with lung cancers who exhibit pathological high-risk factors such as micropapillary, solid, complex glandular, and vascular tumor thrombus, can lead to a higher recurrence rate\cite{Lee2015,Su2020}. Therefore, the accurate identification of stage I lung cancers with pathological high-risk factors is of utmost importance in selecting the appropriate thoracic surgical approach. Examples of high-risk and low-risk nodules are illustrated in Fig. \ref{nodules}. Currently, thoracic surgeons can only decide on the operation mode based on the results of intraoperative frozen pathological sections. Advancing the determination of an appropriate surgical approach before the operation, for instance, based on preoperative CT imaging, remains a significant gap and challenge.
Recently, with the rapid development of artificial intelligence (AI), the application of deep learning models in lung CT diagnosis is becoming increasingly widespread. The majority of research focuses on lung nodule segmentation \cite{Xu2021,Wu2020}, differentiation of benign and malignant nodules \cite{Wang2022,Shen2019}, and recognition of nodule progression \cite{Zhao2023,Wang2020}. For most of the benign and malignant classification tasks, their goal is to detect high risk people from screened samples\cite{Kuang2020,Venkadesh2021}. In addition to that, a number of scholars were interested in identifying pre-invasive and invasive nodules among malignant nodules\cite{Shen2021}. %This can assist in determining which patients require follow-ups and which patients need to undergo surgeries. 
A latest study by Zhou et al., (2023) have built an online platform, which can not only predict benign and malignant, pre-invasive and invasive nodules, but also can classify the invasive nodules into three different grade risk according to the latest IASLC grading system \cite{zhou2023ensemble}.  However, to the best of our knowledge, there is no research on the identification and classification of the pathological high-risk factors of lung cancer based on preopearitve CT images. This may be due to two important reasons. First, obtaining and annotating pulmonary nodules on CT images with precise pathological high-risk factor features is extremely difficult. Second, the features of such nodules are more challenging to capture, necessitating the development of more sophisticated models for prediction.

To this end, we are motivated to develop a novel classification method that can accurately identify the pathological high-risk factors in stage I lung cancer, since it is extremely important for the selection of thoracic surgical mode. Although current lung nodule classification methods have yielded encouraging result, some problems still exit. First, from a clinical perspective, nearly all these methods focus on benign/malignant screening and pre-invasive/invasive classification, with few research efforts directed towards pathological high-risk factor prediction. This limitation restricts their clinical application to some extent. Second, technically, most methods require 3D patches as model input, rendering the training process time-consuming and inefficient.  Finally, for complex lung lesions, such as the discrimination of pathological high-risk factors, existing algorithms still have certain deficiencies in handling large amounts of data and extracting intricate features. To overcome the aforementioned problems in current research, we investigated the classification of pathological high-risk factors for lung cancer in this study. A graph convolutional network with both channel and spatial attention module is proposed to enhance the extraction of characteristic information from pulmonary nodules and improve the training efficiency. In summary, our contributions are as follows:

\begin{itemize}
    \item We propose a graph convolutional neural network that integrates an attention-enhanced feature extractor to utilize the spatial information of pulmonary nodules with pathological high-risk factors.
    \item We develop a novel graph construction method that leverages the nodule's slice-level positional information with GCN, enabling a more effective and less time-consuming model training.
    \item Compared with the benchmarks models, the proposed method has shown a 9.42\% improvement in terms of AUC value. 
\end{itemize}

\section{Related Work}
In this section, we summarize the related research from two aspects, they are respectively, pulmonary nodule classification and GCN model application. 

\subsection{Pulmonary Nodule Classification}
Prior studies for the diagnosis of pulmonary nodules mainly relied on radiomic models, which involved extracting thousands of tumor-related features to quantify various image characteristics of lung tumors, including morphological, texture, boundary, and intensity features \cite{wu2020ct,hu2020non}. %Notably, Wu et al. \cite{wu2020diagnosis} demonstrated the significance of distinguishing ground-glass and solid CT radiomic features of part-solid nodules for diagnosing the invasiveness of lung adenocarcinoma.
Recently, convolutional neural networks (CNNs) have gained increasing attention for their ability to extract features directly from CT images \cite{coudray2018classification,zhao20183d}. For instance, Bonavita et al. proposed a 3D CNN model for evaluating nodule malignancy \cite{Bonavita2020}. Furthermore, attention mechanisms, such as the convolutional
block attention module (CBAM) \cite{woo2018cbam}, have shown promising performance for enhancing deep learning models' accuracy, prompting researchers to explore their application in pulmonary nodule classification. For example, Sun et al. introduced an attention-embedded complementary-stream convolutional neural network (AECS-CNN) capable of extracting contextual features at different scales \cite{SunLing2021}.

Despite these advancements, the majority of research on pulmonary nodule discrimination has focused on classifying benign and malignant nodules or pre-invasive and invasive nodules, with only a few studies addressing the recognition of the subtype of invasive nodules \cite{zhou2023ensemble}. Surprisingly, there is currently no existing research on the classification of pathological high-risk factors based on preoperative CT images, an important factor influencing surgeons' choice of an appropriate surgical approach. This gap in knowledge has inspired us to propose a novel method to predict pathological high-risk factors in pulmonary nodules.

\subsection{GCN Model Application}
The graph convolutional network is an extension of CNN that moves from the Euclidean domain to the graph domain and is extensively used for learning from graph data \cite{shuman2013emerging}. With the ability to address modeling challenges associated with non-Euclidean spatial data and extract essential spatial features, GCN has shown promising prospects across different domains. For instance, Yao et al. introduced a unique approach, constructing a comprehensive graph from an entire corpus and employing a GCN model for text classification \cite{yao2019graph}. Additionally, Choi et al. proposed a GCN-based system that directly estimates the 3D coordinates of human mesh vertices from 2D human pose data \cite{choi2020pose2mesh}.

\begin{figure*}[ht]
\centering
\includegraphics[width=\textwidth]{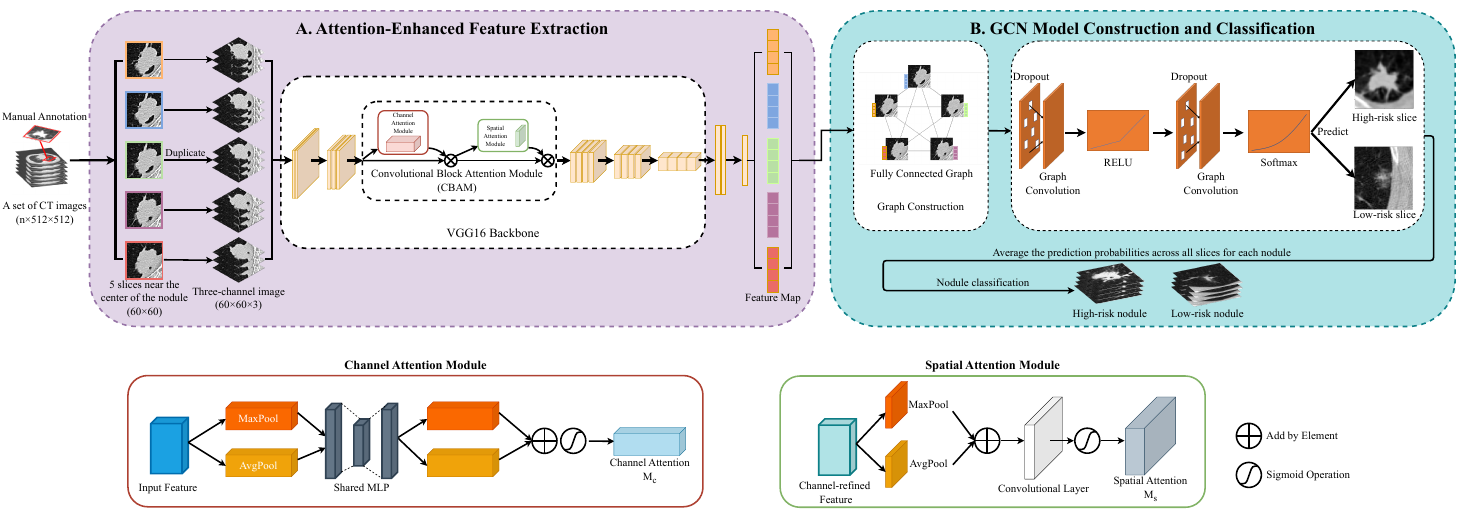}
\caption{The structure of the proposed AE-GCN network. Module A represents the CBAM feature extractor, utilizing both channel and spatial attention mechanism on a pre-trained VGG (a zoom-in view of the attention mechanism is provided below). Module B encompasses both the graph construction and GCN classification. The final nodule-level prediction is obtained by averaging the slice-level probabilities.}
\label{pipeline}
\end{figure*}

For medical imaging analysis, the significance of GCN lies in its ability to capture and incorporate both local and global interactions among image elements during the learning process. For automated anatomical labeling of coronary arteries, Yang et al. introduced the conditional partial-residual graph convolutional network (CPR-GCN), which takes both position and CT image into consideration \cite{yang2020cpr}. Furthermore, Meng et al. combined CNN and GCN to enhance the accuracy of boundary regression in biomedical image segmentation \cite{meng2020cnn}.

Despite GCN has been proven to have considerable advancements in various fields, its application in lung nodule classification remains relatively limited. For instance, Adnan et al. utilized GCN for feature extraction from significant patches on the whole-slide images (WSIs) for subtypes classification of lung cancer \cite{Adnan2020}. However, their approach focused solely on feature extraction and overlooked the inherent spatial structure between each slice of the nodule. Recognizing that the sequence order of the nodule slices automatically forms a graph-like structure, we are motivated to leverage GCN to process extracted features for classification task.

\section{Methodology}
In this section, we introduce the proposed Attention-Enhanced Graph Convolutional Network (AE-GCN) model for pathological high-risk factor identification in lung cancer. Leveraging attention mechanisms for feature extraction and the GCN for classification, our method is good at adaptively extracting deep features from nodules while capitalizing on their inherent spatial structure. An overview of our method is provided in Fig. \ref{pipeline}. Specifically,
let $c^i_k$ be the $i$th slice of a nodule $N_k$, where $k$ index the nodule number. Then the total input slice of node $N_{k}$ is denoted as $C_k=\{c_k^1,...,c_k^{n_k}\}$, with ${n_k}$ denoting the total slice number of nodule $N_{k}$. Our goal is to build a deep neural network $G$ that predicts $Y_k$ given $C_k$: 
$$
P(Y_k=1|C_k)=G(C_k)
$$
where $Y_k \in \{0,1\}$ denotes whether $N_k$ is of pathological high-risk facotrs.

\subsection{Attention-Enhanced Feature Extraction}

For the proposed AE-GCN, a pretrained VGG model is utilized as the backbone for feature extraction. To increase the model performance, we employ a CBAM \cite{woo2018cbam} within the second block of the VGG architecture (See module A in Fig. \ref{pipeline}). This module employs both the channel-wise attention, which assesses the significance of different feature channels, and the spatial attention that identifies vital regions within the feature map. This attentional design refines the model by facilitating the extraction of key features, specifically focusing on the Region of Interest (ROI) within lung tissue. Then, we train the attention-enhanced VGG model, denoted by $V$, on the training set and optimize it on the validation set, thereby tailoring it to effectively extract deep features from pulmonary nodules.

After obtaining $V$, we can extract the corresponding slice-level feature vector $x_k^i \in \mathbb{R}^{1\times D}$ for $c_k^i$ as follows:
\begin{equation}
x_k^i = \text{SecondLastLayer}(V(c_k^i))
\end{equation}

Subsequently, the nodule-level representation or feature map of the nodule $N_k$ can be defined as $X_k = [x_k^1,...,x_k^{n_k}] \in \mathbb{R}^{n_k \times D}$. In our configuration, the feature dimension $D$ is set to be 512.
\subsection{Graph Construction}
To construct the graph, we compile the feature maps of all nodules in the training set into a matrix $X=[X_k]\in \mathbb{R}^{N\times D}$, where $N=\sum{n_k}$ yields the total nodule count. This matrix encapsulates the high-dimensional semantic features of the nodules, with each row corresponding to a specific slice $c_k^i$ of a nodule $N_{k}$. We extract $x_k^i$ to represent the attributes of each slice, and the spatial structure of these slices is then used to establish their connectivity.

For each nodule, the information extracted varies with the number of selected slices, ${n_k}$. We employ two selection strategies for constructing graph-structured data: a fixed approach and a comprehensive approach. For the fixed strategy, we center around the middle slice and uniformly select two slices on either side, thus, ${n_k}=5$. For the comprehensive strategy, we include all CT slices containing the nodule, hence ${n_k}$=SliceNumber$(N_k)$.

\begin{figure}[htbp]
\centering
\includegraphics[width=\columnwidth]{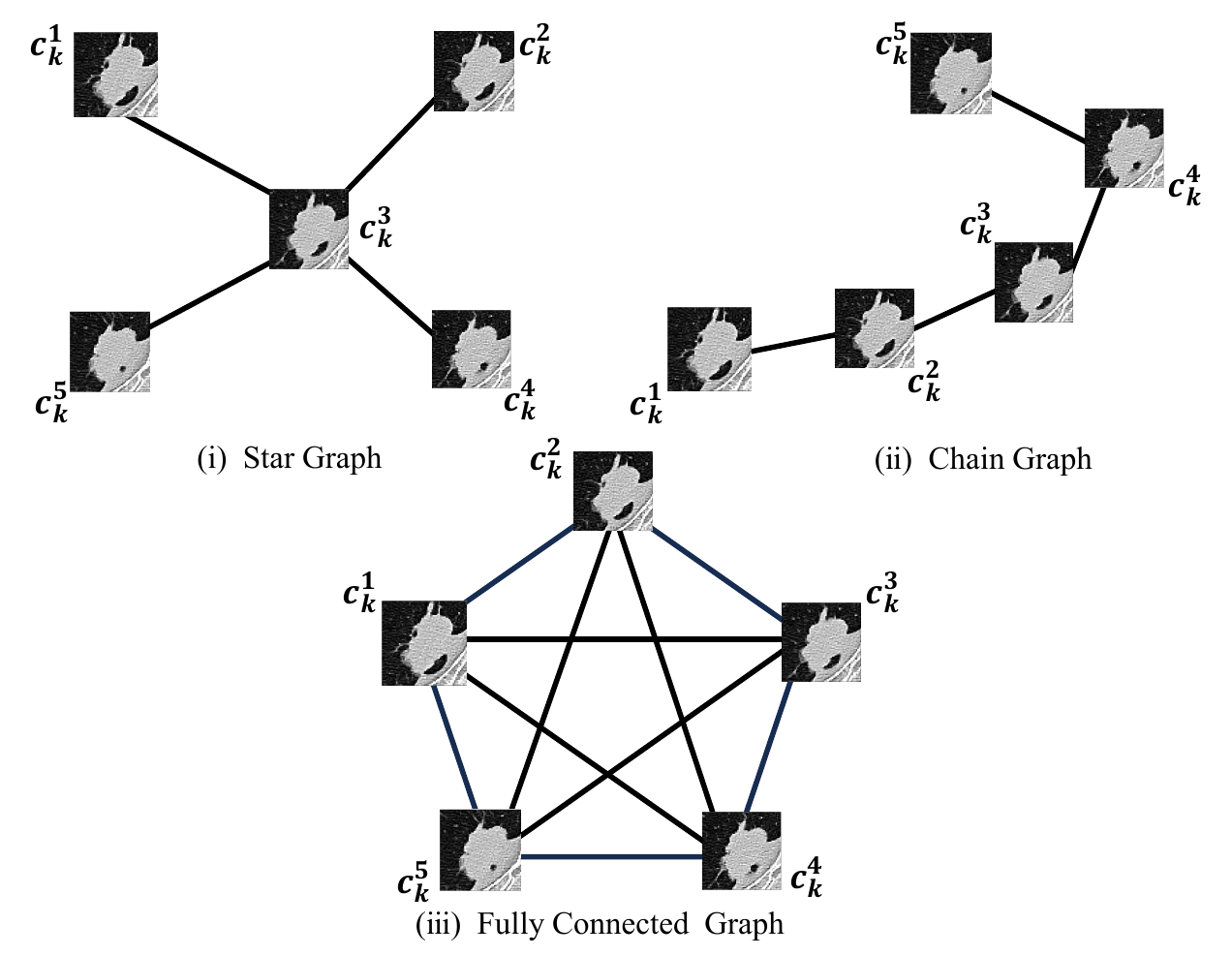}
\caption{Illustration of three graph construction methods: star graph, chain graph, and fully connected graph.}
\label{graphs}
\end{figure}

For graph construction, we evaluate three distinct methods: star graph, chain graph, and fully connected graph. The distinction among them lies in the different approaches they take to model relationships between slices within a nodule. These varying connective structures facilitate different feature representations, which can significantly influence the effectiveness and performance of the proposed model. For demonstrative purposes, we use ${n_k}=5$ as an example and illustrate each method in Fig. \ref{graphs}.

(1) A Star Graph connects each node from the central slice to all other slices, thereby establishing edges between the central node and all others. 

(2) A Chain Graph links each node from the bottom slice to the top slice, creating edges only between adjacent slices. 

(3) A Fully Connected Graph interconnects all slices within each nodule, constructing edges between any two nodes.

Following the chosen method, we generate the adjacency matrix $A$, where $A_{ij}=1$ implies an edge between nodes $c^i_k$ and $c^j_k$ and $A_{ij}=0$ otherwise. Using the adjacency matrix, we successfully transform CT image data into graph-structured representation data, enabling the extraction of spatial feature information for the GCN model.

\subsection{GCN Classification}

After obtaining the feature map $X$ and the adjacency matrix $A$, we employ the structure of the GCN model proposed by Kipf et al\cite{Kipf2016} for classification. This model leverages spectral graph convolutions and its structure is illustrated as module B in Fig. \ref{pipeline}. The specific formula for graph convolution is:
\begin{equation}
H^{(l+1)} = \sigma(\widetilde{D}^{-1/2} \widetilde{A} \widetilde{D}^{-1/2} H^{(l)} W^{(l)})
\end{equation}
with $\widetilde{D}_{ii} = \sum_{j}{\widetilde{A}_{ij}}$. In this context, $\widetilde{A} = A+I_N$ denotes the adjacency matrix with added self-connections, and $I_N$ is the identity matrix. $\widetilde{D}^{-1/2} \widetilde{A} \widetilde{D}^{-1/2}$ represents the Laplacian matrix of the graph. $W^{(l)}$ is a layer-specific trainable weight matrix, $\sigma(\cdot)$ is an activation function, and $H^{(l)}\in \mathbb{R}^{N \times D}$ is the feature matrix of all nodules in the $l_{th}$ layer, where $H^{(0)} = X$.

The proposed GCN model is composed of two dropout layers and two GCN layers. The first GCN layer, featuring 32 hidden dimensions, uses a Leaky ReLU activation function. Subsequently, the second GCN layer, with two dimensions corresponding to our binary classification task, provides the final probabilistic prediction $P$ obtained by applying a softmax activation function. Defining $\hat{A}=\widetilde{D}^{-1/2} \widetilde{A} \widetilde{D}^{-1/2}$, we can simplify the model as follows:
\begin{equation}
P =H^{(2)}=\text{softmax}(\hat{A}\text{ReLu}(\hat{A}XW^{(0)}) W^{(1)})
\end{equation}

Consequently, we have constructed an attention-enhanced GCN model for the identification of high-risk factors in lung cancer, highlighting its potential to detect high-risk pulmonary nodules. 

\begin{table*}[h]
\centering
\caption{ Performance Comparison of Graph Construction Methods and Patch Number in AE-GCN }\label{tab_var}
% \usepackage{multirow}
\begin{tabular*}{\textwidth}{@{\extracolsep{\fill}}llccccccc}
\hline
Graph Construction & Patch Number & AUC    & Acc    & Spe    & Sen    & PPV    & NPV    & F1     \\ \hline
Fully Connected    & All          & 0.9524 & 0.9355 & 0.9104 & 0.8462 & 0.7857 & 0.9385 & 0.8148 \\
Fully Connected    & 5            & 0.9432 & 0.9451 & 0.9077 & 0.8846 & 0.7931 & 0.9516 & 0.8364 \\
Chain              & All          & 0.9420 & 0.9355 & 0.9552 & 0.8846 & 0.8846 & 0.9552 & 0.8846 \\
Chain              & 5            & 0.9314 & 0.9011 & 0.9692 & 0.7308 & 0.9048 & 0.9000 & 0.8085 \\
Star               & All          & 0.9351 & 0.9032 & 0.9403 & 0.8077 & 0.8400 & 0.9265 & 0.8235 \\
Star               & 5            & 0.9207 & 0.9011 & 0.8308 & 0.8077 & 0.6563 & 0.9153 & 0.7241 \\ \hline

\end{tabular*}
\end{table*}

\begin{table*}[h]
\centering
\caption{ Comparison of various methods in high-risk factor prediction in lung cancer}\label{tab_com}
% \usepackage{multirow}
\begin{tabular*}{\textwidth}{@{\extracolsep{\fill}}lccccccc}
\hline
Model     & AUC             & Acc             & Spe             & Sen             & PPV             & NPV             & F1              \\ \hline
AE-GCN    & \textbf{0.9524} & \textbf{0.9355} & \textbf{0.9104} & 0.8462          & \textbf{0.7857} & \textbf{0.9385} & \textbf{0.8148} \\
Resnet    & 0.7738          & 0.7634          & 0.8507          & 0.5385          & 0.5833          & 0.8261          & 0.5600          \\
DenseNet  & 0.7308          & 0.6989          & 0.7164          & 0.6538          & 0.4722          & 0.8421          & 0.5484          \\
3D Resnet & 0.7692          & 0.7957          & \textbf{0.9104} & 0.5000          & 0.6842          & 0.8243          & 0.5778          \\
ViT       & 0.8186          & 0.7312          & 0.6716          & \textbf{0.8846} & 0.5111          & 0.9375          & 0.6479          \\
ViT-GCN   & 0.8582          & 0.7849          & 0.8060          & 0.6923          & 0.5806          & 0.8710          & 0.6316          \\ \hline
\end{tabular*}
\end{table*}
\section{Experiments}

\subsection{Dataset}

The dataset utilized in this study was obtained from two hospitals, including a total of 483 pulmonary nodules collected from 426 patients. Among these nodules, 139 have been identified with pathological high-risk factors, such as micropapillary, solid, complex glandular and vascular tumor thrombus. To ensure an unbiased comparison, we randomly partitioned the patients into three distinct subsets: training, validation, and testing, with an approximate ratio of 6:2:2. The center coordinates (i.e., $X, Y, Z$) of each pulmonary nodule were annotated by an experienced thoracic radiologists and their pathological sections were independently reviewed by a professional pathologist to determine whether there are high-risk factors. %Our objective is to utilize the proposed model to accurately discern these high-risk nodules.

The raw CT scans were in Digital Imaging and Communications in Medicine (DICOM) format, with each slice possessing a resolution of 512$\times$512. In the preprocess procedure, we first converted the DICOM data into Hounsfield Unit (HU) image matrices trimmed to [-1400, 400], which allows our focus on the lung tissues. Subsequently, we normalized the CT values to fall within the range [0, 1] and then centered them around 0 by subtracting the mean value. Finally, the processed images of each nodule were cropped into $n_k$ patches of 60$\times$60. For data augmentation, we applied various image transformations, such as flipping, rotation, and swapping. Given the single-channel nature of grayscale CT images, we replicated these into a three-channel format, making them suitable for use as input to the proposed CNN feature extractor.

\subsection{Experimental Settings}
The feature extraction module was fine-tuned with an initial learning rate of 0.001. If no accuracy improvement was detected after 20 epochs, we reduced the learning rate by 50\%. The total training process was set to 50 epochs, with Adam as the optimizer. For the GCN classification module, the model was empirically configured with a learning rate of 0.0001, a dropout rate of 0.3, and a training duration of 200 epochs. Across both modules, we employed the cross-entropy loss as our loss function to address the binary classification problem. Each model was fitted on the training set, and the model that demonstrated the highest accuracy on the validation set was selected for further evaluation.

\subsection{Evaluation Metrics}
We employ a range of common metrics to assess the model performance. These include the Area Under the ROC Curve (AUC), accuracy (Acc), sensitivity (Sen), specificity (Spe), positive predictive value (PPV), negative predictive value (NPV), and F1-Score. In the GCN classification module, our model makes slice-level predictions. Recognizing that a single nodule consists of multiple slices, we average the prediction probabilities across all slices for each nodule to obtain a comprehensive nodule-level prediction. Performance metrics are then calculated based on these nodule-level predictions.

\subsection{Model Variations}
To assess the influence of different components within the AE-GCN model, we investigate several modifications, focusing on graph construction methods and the number of slices used. First, we investigated the influence of three distinct graph construction methods: star graph, chain graph, and fully connected graph. Second, we explored the impact of data volume by employing two strategies for the number of slices (i.e., ${n_k}$): a fixed strategy with ${n_k}=5$, and a comprehensive strategy including all slices for each nodule with ${n_k}= \text{SliceNumber}(N_k)$.

The evaluation results are summarized in Table \ref{tab_var}. For model performance evaluation, we used the AUC value as our primary metric and focused on nodular-level metrics. Overall, the fully connected graph outperformed other structures, regardless of the value of ${n_k}$. This can be attributed to its capability to better utilize spatial features within the nodules. Furthermore, the comprehensive strategy consistently led to higher AUC values, underlining the benefit of incorporating more information. These results suggest that a more complex graph structure, together with an increased volume of information, can enhance our model's effectiveness. Consequently, we chose a fully connected graph with a comprehensive strategy as the optimal model for further analysis.
\begin{table*}[h]
\centering
\caption{ Ablation Study on attention mechanisms}\label{tab_abl}
\begin{tabular*}{\textwidth}{@{\extracolsep{\fill}}llccccccc}
\hline
Channel Attention & Spatial Attention & AUC             & Acc             & Spe             & Sen             & PPV             & NPV             & F1              \\ \hline
Yes               & Yes               & \textbf{0.9524} & \textbf{0.9355} & \textbf{0.9104} & \textbf{0.8462} & \textbf{0.7857} & \textbf{0.9385} & \textbf{0.8148} \\
Yes               & No                & 0.8123          & 0.7742          & 0.7910          & 0.6923          & 0.5625          & 0.8689          & 0.6207          \\
No                & Yes               & 0.8352          & 0.7742          & 0.8209          & 0.6923          & 0.6000          & 0.8730          & 0.6429          \\
No                & No                & 0.7732          & 0.7634          & 0.7910          & 0.6923          & 0.5625          & 0.8689          & 0.6207          \\ \hline
\end{tabular*}
\end{table*}

\subsection{Comparison of Different Methods}
We compared our proposed model on the testing set against several models that have been extensively studied in previous reports. For end-to-end models, we utilized 2D CNN models such as DenseNet \cite{huang2017densely} and ResNet18 \cite{ResNet}, as well as the recent state-of-the-art Vision Transformer (ViT) \cite{dosovitskiy2021an}, and a 3D ResNet18 model \cite{chen2019med3d}. For hybrid models, we also evaluated a GCN model with ViT serving as the feature extractor (ViT-GCN). This was done to validate the effectiveness of the attention mechanism in comparison to the transformer structure for leveraging contextual information in feature extraction. The results of these comparisons are presented in Table \ref{tab_com}. As we can see, the proposed model outperformed others across most metrics by a considerable margin, with a 9.42\% increase for AUC and 13.98\% for Acc. This result demonstrated the effectiveness of combining attention-based feature extraction with GCN classification. Specifically, the advantage of the CBAM over the transformer is evident from the superior performance of AE-GCN compared to ViT-GCN. Interestingly, both hybrid models showed comparably higher performance, which can be attributed to the effectiveness of incorporating spatial information via GCN. Meanwhile, context-aware models (i.e., AE-GCN, ViT, and ViT-GCN) consistently surpassed other models in performance, even outperforming 3D Resnet, which is more complex in terms of parameter count. This indicates the importance of employing a more nuanced feature extractor. In addition to performance comparisons, we also conducted a brief analysis of the training time for the models.The results highlight the efficiency of the proposed AE-GCN model, which takes, on average, less than one second per epoch to complete. In contrast, the traditional 2D CNN model requires over a dozen seconds per epoch, and the 3D CNN model is even more time-consuming, exceeding one minute per epoch. This efficiency in training further underscores the practicality of our approach.

\subsection{Ablation Study}
To assess the individual and collective contributions of the channel and spatial attention modules within the CBAM module, we conducted an ablation study using the testing set. This examination scrutinized the impact of including or excluding these attention mechanisms, with the outcomes shown in Table \ref{tab_abl}. The results reveal that both the channel and spatial attention yielded significant improvements in AUC: 11.72\% for channel attention, 14.01\% for spatial attention, and an even more pronounced 17.92\% when combined. These findings underscore the modules' capability to discern and capture important features, particularly when employed together.

\section{Conclusion}

Accurate identification of early-stage lung cancers with pathological high-risk factors is crucial for surgical mode selection. At present, thoracic surgeons usually rely on intraoperative frozen pathological section results to guide their choice of surgical approach. However, there remains a significant gap and challenge in determining the suitable surgical mode before the operation, such as based on preoperative CT images. In this paper, we propose a AE-GCN model, a novel approach that synergizes attentional features with spatial information to address this challenging issue. We develop a novel graph construction method that leverages the nodule's positional information with GCN, which enables us to train the model more effectively and less time-consuming. Experimental results demonstrated that the proposed model outperforms many of the previous benchmark methods, yielding an increase of 9.42\% in terms of AUC value. By utilizing the proposed AE-GCN model, thoracic surgeons are given an opportunity to evaluate the high-risk probability of a pulmonary nodule solely based on CT images, which could also assist surgeons in making preoperative surgical plans.

\bibliographystyle{IEEEtran}
\bibliography{bibliography}

\end{document}